\documentclass[aps,showpacs,twocolumn]{revtex4}%
\usepackage{amsfonts}
\usepackage{amsmath}
\usepackage{amssymb}
\usepackage{graphicx}%
\begin{document}
\title{Edge states and the integer quantum Hall conductance in spin-chiral
ferromagnetic kagom\'{e} lattice}
\author{Zhigang Wang and Ping Zhang}
\affiliation{Institute of Applied Physics and Computational Mathematics, P.O. Box 8009,
Beijing 100088, P.R. China}

\begin{abstract}
We investigate the chiral edge states in the two-dimensional ferromagntic
kagom\'{e} lattice with spin anisotropies included. The system is periodic in
the $x$ direction but has two edges in the $y$ direction. The Harper equation
for solving the energies of edge states is derived. We find that there are two
edge states in each bulk energy gap, corresponding to two zero points of the
Bloch function on the complex-energy Riemann surface (RS). The edge-state
energy loops parametrized by the momentum $k_{x}$ cross the holes of the RS.
When the Fermi energy lies in the bulk energy gap, the quantized Hall
conductance is given by the winding number of the edge states across the
holes, which reads as $\sigma_{xy}^{\text{edge}}$=$-\frac{e^{2}}{h}%
$sgn$\left(  \sin\phi\right)  $, where $\phi$ is the spin chiral parameter
(see text). This result keeps consistent with that based on the topological
bulk theory.

\end{abstract}
\maketitle

Recently the quantum transport of electrons in spin-orbit coupled
\cite{Jungwirth,Fang,Yao} or spin-chiral ferromagnetic systems
\cite{Matl,Chun,Ye,Tag} has been a focus of intense interest in condensed
matter physics. One typical spin-chiral ferromagnetic system is represented by
pyrochlore compounds $R_{2}$Mo$_{2}$O$_{7}$ ($R$=Nd, Sm, Gd), in which the
spin configuration is noncoplanar and the spin chirality appears. As a
consequence, the quantum transport of electrons, especially the transverse
conductivity $\sigma_{xy}$, is expected to be affected by the presence of spin
chirality. Ohgushi et al. \cite{Ohgushi} have first pointed out that the
chiral spin state can be realized by the introduction of spin anisotropy in an
\textit{ordered} spin system on the two-dimensional (2D) kagom\'{e} lattice,
which is the cross section of the pyrochlore lattice perpendicular to the
$(1,1,1)$ direction \cite{Ramirez}. In this case, it has been shown in the
topological bulk theory \cite{Ohgushi,Wang2007} that the presence of chiral
spin state may induce gauge-invariant nonzero Chern number, thus resulting in
a quantized Hall effect in insulating state.

In this paper we turn to study the 2D kagom\'{e} lattice with two edges,
which, as will be shown below, displays two \emph{chiral} (instead of
\textit{nonchiral}) edge states localized near the sample boundaries. Closely
following the topological edge theory established in last decade
\cite{Hatsugai1,Hatsugai2, Hatsugai3}, we first derive the transfer matrix
(namely, the Harper equation \cite{Harper, Hosfstadter,Wannier}) for solving
the energies of the edge states. Although the transfer matrix elements are no
longer polynomials of the energy $\epsilon$ for a fixed spin chiral parameter
$\phi$ (except for special cases $\phi$=$0,\pm\pi/2,\pi$), we find that the
transfer matrix method is also applicable in this system. Then by numerical
calculation, we find that there are two edge states in each bulk energy gap,
corresponding to two zero points of the Bloch function on the complex-energy
Riemann surface (RS). Remarkably different from the case of a square lattice
under a magnetic field \cite{Hatsugai1}, the edge-state energy loop moves
\emph{across} (\emph{not} \emph{around}) the holes in the RS in the present
model. The two edge-state energy loops lying in the same energy gap are
tangent at one point and their appearance shows a \textquotedblleft$\infty
$\textquotedblright\ structure. Furthermore, we obtain that when the Fermi
energy lies in the bulk energy gap, the quantum Hall conductance given by the
winding number of the edge state can be written as $\sigma_{xy}^{\text{edge}}%
$=$-\frac{e^{2}}{h}$sgn$\left(  \sin\phi\right)  $. This result based on the
topological edge theory keeps consistant with that based on the topological
bulk theory \cite{Ohgushi,Wang2007}.

Following Ref. \cite{Ohgushi}, we consider the double-exchange ferromagnet on
the kagom\'{e} lattice schematically shown in the left panel in Fig. 1
\cite{Ohgushi, Wang2007}. Here the triangle is the one face of the
tetrahedron, and the easy axis of the spin anisotropy points to the center of
each tetrahedron and has an out-of-plane component. In this situation the
three local spins on sites A, B, and C in the left panel in Fig. 1 have
different directions and the spin chirality emerges. The effective Hamiltonian
for the hopping electrons strongly Hund-coupled to these localized spins is
given by $H=\sum_{NN}t_{ij}^{eff}c_{i}^{\dag}c_{j}$ with $t_{ij}%
^{eff}=t\langle\chi_{i}|\chi_{j}\rangle=te^{ia_{ij}}\cos\frac{\vartheta_{ij}%
}{2}$. Here the spin wave function $|\chi_{i}\rangle$ is explicitly given by
$|\chi_{i}\rangle=\left[  \cos\frac{\vartheta_{i}}{2},\text{ }e^{i\phi_{i}%
}\sin\frac{\vartheta_{i}}{2}\right]  ^{\text{T}}$, where the polar coordinates
are pinned by the local spins, i.e., $\langle\chi_{i}|\mathbf{S}_{i}|\chi
_{i}\rangle=\frac{1}{2}\left(  \sin\vartheta_{i}\cos\phi_{i},\text{ }%
\sin\vartheta_{i}\sin\phi_{i},\text{ }\cos\vartheta_{i}\right)  $.
$\vartheta_{ij}$ is the angle between the two spins $\mathbf{S}_{i}$ and
$\mathbf{S}_{j}$. The phase factor $a_{ij}$ can be regarded as the gauge
vector potential $a_{\mu}(\mathbf{r})$, and the corresponding gauge flux is
related to scalar spin chirality $\chi_{ijk}$=$\mathbf{S}_{i}{\small \cdot
}(\mathbf{S}_{j}{\small \times}\mathbf{S}_{k})$ \cite{Laughlin2}. In periodic
crystal lattices, the nonvanishing of the gauge flux relies on the multiband
structure with each band being characterized by a Chern number
\cite{Thouless,Shindou}. Following Ref. \cite{Ohgushi,Wang2007} we set the
flux originated from the spin chirality per triangle (see Fig. 1) as $\phi$,
which satisfies $e^{i\phi}=e^{i(a_{AB}+a_{BC}+a_{CA})}$. The flux penetrating
one hexagon is determined as $-2\phi$. We take the gauge, in which the phase
of $t_{ij}^{eff}$ is the same for all the nearest-neighbor pairs with the
direction shown by the arrows in the left panel of Fig. 1. It should be
pointed out that the net flux through a unit cell vanishes due to the
cancelation of the contribution of the two triangles and a hexagon. Also noted
is that the time-reversal symmetry is broken except for cases of $\phi$%
=$0$,$\pi$. In the following we change notation $i\rightarrow(lms)$, where
$\left(  lm\right)  $ label the kagom\'{e} unit cell and $s$ denote the sites
A, B and C in this cell. The size of the unit cell is set to be unity
throughout this paper.\begin{figure}[ptb]
\begin{center}
\includegraphics[width=1.0\linewidth]{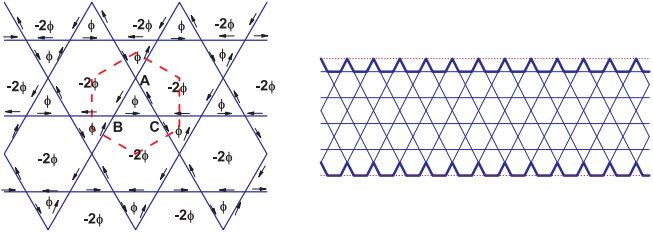}
\end{center}
\caption{(Color online) Left panel: Two dimensional spin-chiral ferromagnetic
kagom\'{e} lattice. The dashed line represents the Wigner-Seitz unit cell,
which contains three independent sites (A, B, C). It is assumed that each site
has a different spin anisotropy axis. The arrows means the sign of the phase
of the transfer integral $t_{ij}$. Right panel: The 2D kagom\'{e} lattice
system with edges along the $y$ direction.}%
\end{figure}

Now we suppose that the system is periodic in the $x$ direction but has two
edges in the $y$ direction (see the right panel of Fig. 1). Since the system
is periodic in the $x$ direction, we can use a momentum representation of the
electron operator
\begin{equation}
c_{(lms)}=\frac{1}{\sqrt{L_{x}}}\sum_{k_{x}}e^{ik_{x}X_{(lms)}}\gamma
_{ms}(k_{x}),
\end{equation}
where $\left(  X_{(lms)},Y_{(lms)}\right)  $ are the coordinate of the site
$s$ in the unit cell $(lm)$ and $k_{x}$ is the momentum along the $x$
direction. Let us consider the one-particle state $|\Psi(k_{x})\rangle$%
=$\sum_{ms}$ $\Psi_{ms}(k_{x})\gamma_{ms}^{\dag}(k_{x})|0\rangle$. Inserting
it into the Schr\"{o}dinger equation $H|\Psi\rangle$=$\epsilon|\Psi\rangle$,
we can easily obtain the following three eigenvalue equations for sites A, B,
and C,%
\begin{align}
\epsilon\Psi_{mA}  &  =e^{-i\frac{\phi}{3}}\left[  e^{i\frac{1}{4}k_{x}}%
\Psi_{mB}+e^{-i\frac{1}{4}k_{x}}\Psi_{(m+1)B}\right] \nonumber\\
&  +e^{i\frac{\phi}{3}}\left[  e^{-i\frac{1}{4}k_{x}}\Psi_{mC}+e^{i\frac{1}%
{4}k_{x}}\Psi_{(m+1)C}\right]  ,\nonumber\\
\epsilon\Psi_{mB}  &  =2e^{-i\frac{\phi}{3}}\cos\left(  \frac{k_{x}}%
{2}\right)  \Psi_{mC}\nonumber\\
&  +e^{i\frac{\phi}{3}}\left[  e^{-i\frac{1}{4}k_{x}}\Psi_{mA}+e^{i\frac{1}%
{4}k_{x}}\Psi_{(m-1)A}\right]  ,\nonumber\\
\epsilon\Psi_{mC}  &  =e^{-i\frac{\phi}{3}}\left[  e^{i\frac{1}{4}k_{x}}%
\Psi_{mA}+e^{-i\frac{1}{4}k_{x}}\Psi_{(m-1)A}\right] \label{equations}\\
&  +2e^{i\frac{\phi}{3}}\cos\left(  \frac{k_{x}}{2}\right)  \Psi
_{mB}.\nonumber
\end{align}
Eliminating the B- and C-sublattice sites, we obtain the difference equation,%
\begin{align}
&  \left[  \epsilon^{3}-4\epsilon\left(  1+\cos^{2}\left(  \frac{k_{x}}%
{2}\right)  \right)  -8\cos^{2}(\frac{k_{x}}{2})\cos\phi\right]  \Psi
_{mA}\nonumber\\
&  =2\cos\left(  \frac{k_{x}}{2}\right)  \left(  \epsilon+2\cos\phi\right)
\left[  \Psi_{(m+1)A}+\Psi_{(m-1)A}\right]  . \label{eq}%
\end{align}
This is the Harper equation \cite{Harper, Hosfstadter}. Equation (\ref{eq})
can be represented in the following matrix form:%
\begin{equation}
\left(
\begin{array}
[c]{c}%
\Psi_{(m+1)A}\\
\Psi_{mA}%
\end{array}
\right)  =\tilde{M}\left(  \epsilon\right)  \left(
\begin{array}
[c]{c}%
\Psi_{mA}\\
\Psi_{(m-1)A}%
\end{array}
\right)  , \label{matrix}%
\end{equation}
where%
\begin{equation}
\tilde{M}\left(  \epsilon\right)  =\left(
\begin{array}
[c]{cc}%
p & -1\\
1 & 0
\end{array}
\right)  \label{m1}%
\end{equation}
and $p$=$\frac{\epsilon\left(  \epsilon^{2}-4\right)  }{2\cos\left(
k_{x}/2\right)  \left(  \epsilon+2\cos\phi\right)  }-2\cos\left(  \frac{k_{x}%
}{2}\right)  $. In the following we do not explicitly write the subscript A in
Eq. (\ref{matrix}). Then we get a reduced transfer matrix linking the two
edges as follows%
\begin{equation}
\left(
\begin{array}
[c]{c}%
\Psi_{L_{y}+1}\\
\Psi_{L_{y}}%
\end{array}
\right)  =M\left(  \epsilon\right)  \left(
\begin{array}
[c]{c}%
\Psi_{1}\\
\Psi_{0}%
\end{array}
\right)  , \label{reducedmatrix}%
\end{equation}
where%
\begin{equation}
M\left(  \epsilon\right)  =\tilde{M}\left(  \epsilon\right)  ^{L_{y}}=\left(
\begin{array}
[c]{cc}%
M_{11}\left(  \epsilon\right)  & M_{12}\left(  \epsilon\right) \\
M_{21}\left(  \epsilon\right)  & M_{22}\left(  \epsilon\right)
\end{array}
\right)  . \label{m}%
\end{equation}
For general $\phi$, which varies in a range between $-\pi$ and $\pi$,
$M_{ij}\left(  \epsilon\right)  $ ($i,j$=$1,2$) are not polynomials of
$\epsilon$. At four special values, i.e., $\phi$=$0,\pm\frac{\pi}{2},\pi$,
however, they can be written as polynomials of $\epsilon$, with the degree of
$2L_{y}$ for $M_{11}$, $2L_{y}-1$ for $M_{12}$ and $M_{21}$, and $2L_{y}-2$
for $M_{22}$. In fact, in the spin-chiral cases of $\phi$=$\pm\frac{\pi}{2}$,
the factor $p$ in $M_{ij}$ is reduced to $p$=$\frac{\epsilon^{2}-4}%
{2\cos\left(  k_{x}/2\right)  }-2\cos\left(  \frac{k_{x}}{2}\right)  $.
Whereas, in the spin-nonchiral cases of $\phi$=$0,\pi$, the factor $p$ in
$M_{ij}$ is reduced to $p$=$\frac{\epsilon\left(  \epsilon\mp2\right)  }%
{2\cos\left(  k_{x}/2\right)  }-2\cos\left(  \frac{k_{x}}{2}\right)  $. All
kinds of solutions from Eq. (\ref{reducedmatrix}) are obtained by different
choices of $\Psi_{0}$ and $\Psi_{1}$.

Now we investigate the energy spectrum of the one-dimensional problem with
special attention to the edge states. The boundary condition of this problem
is
\begin{equation}
\Psi_{L_{y}}=\Psi_{0}=0. \label{boundary}%
\end{equation}
With Eqs. (\ref{reducedmatrix}) and (\ref{m}), one can easily obtain that the
solutions satisfy
\begin{equation}
M_{21}\left(  \epsilon\right)  =0. \label{solution}%
\end{equation}
From%
\[
\left(
\begin{array}
[c]{c}%
\Psi_{L_{y}+1}\\
\Psi_{L_{y}}%
\end{array}
\right)  =\tilde{M}^{2}\left(  \epsilon\right)  \left(
\begin{array}
[c]{c}%
\Psi_{L_{y}-1}\\
\Psi_{L_{y}-2}%
\end{array}
\right)  =M\left(  \epsilon\right)  \left(
\begin{array}
[c]{c}%
\Psi_{1}\\
0
\end{array}
\right)  ,
\]
one can find that
\begin{equation}
\Psi_{L_{y}-1}=-M_{11}\left(  \epsilon\right)  \Psi_{1}. \label{m11}%
\end{equation}
If we use a usual normalized wave function, the state is localized at the
edges as%
\begin{align}
|M_{11}\left(  \epsilon\right)  |  &  \ll1\text{, \ \ \ localized at }%
y\approx1\text{ (down edge),}\label{edge}\\
|M_{11}\left(  \epsilon\right)  |  &  \gg1\text{, \ \ \ localized at }y\approx
L_{y}-1\text{ (up edge).}\nonumber
\end{align}

Because the analytical derivation is very difficult, we now start a numerical
calculation from Eq. (\ref{equations}) and draw in figure 2 the energy
spectrum of the fixed boundary system as a function of $k_{x}$ for three
values of spin-chiral parameter $\phi$. The number of sites A (or B, C) in the
$y$ direction is chosen to be $L_{y}$=$50$. Clearly, one can see that the edge
states occur in the energy gaps or at the band edges. From Fig. 2(b) and (c),
one can clearly observe that in the spin-chiral cases, i.e., $\phi\neq0$ (or
$\pi)$, there are three dispersed energy bands (the shaded areas) with two
edge states (the lines) lying in each energy gap. This feature is different
from that in the case of the square lattice in an external magnetic field
\cite{Hatsugai1}, in which each gap has only one edge state. The reason for
this difference is that the factor $p$ in $M_{ij}(\epsilon)$ is no longer a
linear function of $\epsilon$ in the present case. \begin{figure}[ptb]
\begin{center}
\includegraphics[width=1.0\linewidth]{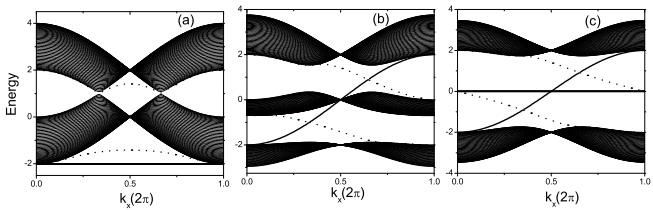}
\end{center}
\caption{Energy spectrum of the two dimensional spin-chiral ferromagnetic
kagom\'{e} lattice system with fixed boundary under different spin-chiral
parameter (a) $\phi$=0, (b) $\phi$=$\pi$/3, and (c) $\phi$=$\pi$/2. The shaded
areas are the energy bands and the lines are the spectrum of the edge states.
The solid and dotted line mean that the edge state is localized near
$y\approx1$ and $y\approx L_{y}-1$, respectively.}%
\end{figure}

Before studying the Hall conductance of this system, we simply review the
winding number \cite{Hatsugai1, Hatsugai3}, which is as well as the Chern
number a well-defined topological quantity. Let us consider the bulk Bloch
function at sites with $y$-coordinate of $L_{y}$. For the Bloch function,
$\Psi_{1}^{(b)}$ and $\Psi_{0}^{(b)}$ compose an eigenvector of $M$ with the
eigenvalue $\rho$,%
\begin{equation}
M\left(  \epsilon\right)  \left(
\begin{array}
[c]{c}%
\Psi_{1}^{(b)}\\
\Psi_{0}^{(b)}%
\end{array}
\right)  =\rho\left(  \epsilon\right)  \left(
\begin{array}
[c]{c}%
\Psi_{1}^{(b)}\\
\Psi_{0}^{(b)}%
\end{array}
\right)  . \label{Bloch}%
\end{equation}
We extend the energy $\epsilon$ to a complex energy to discuss a wave function
of the edge state. Here we use complex variable $z$ for the energy. From Eq.
(\ref{Bloch}) we get%
\begin{equation}
\rho\left(  z\right)  =\frac{1}{2}\left[  \Delta\left(  z\right)
-\sqrt{\Delta^{2}\left(  z\right)  -4}\right]  , \label{p}%
\end{equation}
and
\begin{equation}
\Psi_{L_{y}-1}(z)=-\frac{M_{11}(z)+M_{22}(z)-\sqrt{\Delta^{2}\left(  z\right)
-4}}{-M_{11}(z)+M_{22}(z)+\sqrt{\Delta^{2}\left(  z\right)  -4}}M_{21}(z),
\label{q}%
\end{equation}
where $\Delta\left(  z\right)  =$Tr$\left[  M\left(  z\right)  \right]  $ and
$\Psi_{1}=1$ used. Since the analytic structure of the wave function is
determined by $\omega$=$\sqrt{\Delta^{2}\left(  z\right)  -4}$, we consider
the RS of a hyperelliptic curve $\omega^{2}$=$\Delta^{2}\left(  z\right)  -4$.
To make the analytic structure of $\omega$ to be unique, we have to specify
the brunch cuts which are given by $\Delta^{2}\left(  z\right)  -4\leq0$ at
$\mathfrak{J}z$=$0$. Since this condition also gives the condition for
$\left\vert \rho\right\vert $=$1$, the branch cuts are given by the three
energy bands. Therefore, $\Delta^{2}\left(  z\right)  -4$ can be factorized as%
\[
\omega=\sqrt{\Delta^{2}\left(  z\right)  -4}=\sqrt{%
{\displaystyle\prod\nolimits_{i=1}^{6}}
(z-\lambda_{i})},
\]
where $\lambda_{i}$ denote energies of the band edges. The RS is obtained by
gluing the two Riemann spheres at these branch cuts along the arrows (see Fig.
3). The Riemann spheres are obtained by compactifying the $|z|$=$\infty$
points to one point. After the gluing operation, the surface is topologically
equivalent to the surface shown in Fig. 4. In the present model, the genus of
the RS is $g$=$2$, which is the number of energy gaps. In this way, the wave
function is defined on the genus-$2$ RS $\Sigma_{g=2}(k_{x})$. The branch of
the function is specified as $\Delta^{2}\left(  z\right)  -4>0$ ($z\rightarrow
-\infty$ on the real axis of $R^{+}$). Then if $z$ lies in the $j$th gap from
below on the real axis (notice that there are two real axes), $\alpha\left(
-1\right)  ^{j}\sqrt{\Delta^{2}\left(  z\right)  -4}\geq0$, $z$ (real) on
$R^{\alpha}$ ($\alpha$=$+$, $-$). So, at the energies of the edge states
$\mu_{j}$,
\begin{align}
\sqrt{\Delta^{2}\left(  \mu_{j}\right)  -4}  &  =\alpha\left(  -1\right)
^{j}\left\vert M_{11}(\mu_{j})-M_{22}(\mu_{j})\right\vert \text{ \ }%
\label{mu}\\
&  \left(  \mu_{j}\in R^{\alpha},\alpha=+,-\right)  .\nonumber
\end{align}
In addition, by simple calculation, we can also obtain
\begin{equation}
\Delta\left(  \epsilon\right)  \left\{
\begin{array}
[c]{c}%
\leq-2\text{ \ for\ }j\text{ odd}\\
\geq2\text{ }\ \text{for }j\text{ even}%
\end{array}
\right.  \text{,} \label{delta}%
\end{equation}
where the energy $\epsilon$ (on $R^{\pm}$) is in the $j$th gap. From Eqs.
(\ref{mu}), (\ref{delta}) and (\ref{edge}), we can get that when the zero
point is on the upper sheet of the RS, the edge state is localized at the down
edge; when the zero point is on the lower sheet of the RS, the edge state is
localized at the up edge. \begin{figure}[ptb]
\begin{center}
\includegraphics[width=1.0\linewidth]{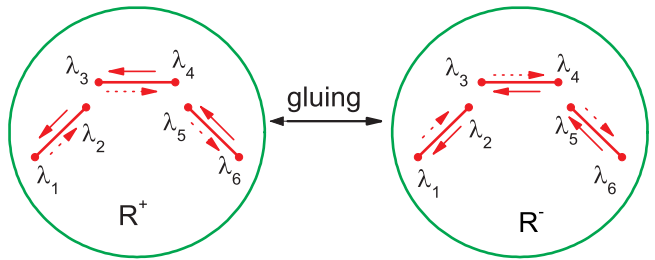}
\end{center}
\caption{(Color online) Two sheets (Riemann spheres) with $3$ cuts which
correspond to the energy bands of the system. The RS of the Bloch function is
obtained by gluing the two spheres along the arrows near the cuts.}%
\end{figure}

In Fig. 4, on the RS, the energy gaps correspond to circles around the holes
of the $\Sigma_{g=2}(k_{x})$ and the energy bands correspond to closed paths
on $\Sigma_{g=2}(k_{x})$. The Bloch function is defined on this surface.
$\Psi_{L_{y}-1}^{(b)}$ has always $2g$=$4$ zero points at the edge-state
energy $\mu_{j}$ ($\Psi_{L_{y}-1}^{(b)}\left(  \mu_{j}\right)  =0$). Since
there are two real axes on the $\Sigma_{g=2}(k_{x})$, there are eight $\mu
_{j}$'s on the RS. However, only one of every two gives a zero of $\Psi
_{L_{y}-1}^{(b)}$.

Changing $k_{x}$ in one period, we can consider a family of $\Sigma
_{g=2}(k_{x})$. $\Sigma_{g=2}(k_{x})$ can be modified by this change yet all
the $\Sigma_{g=2}(k_{x})$ with different $k_{x}$'s are topologically
equivalent if there are stable energy gaps in the two-dimensional spectrum. By
identifying the topologically equivalent $\Sigma_{g=2}(k_{x})$, we can observe
that the $\mu_{j}\left(  k_{x}\right)  $ moves \emph{across} the holes and
forms an oriented loop $C\left(  \mu_{j}\right)  $. Note that the present case
is prominently different from the previous case of the square lattice under an
external magnetic field \cite{Hatsugai1}, in which $\mu_{j}$ moves
\emph{around} the holes. The two edge-state energy loops $C$'s in the same
energy gap are tangent at one point and their appearance shows a
\textquotedblleft$\infty$\textquotedblright\ structure, as shown in Fig.
4.\begin{figure}[ptb]
\begin{center}
\includegraphics[width=1.0\linewidth]{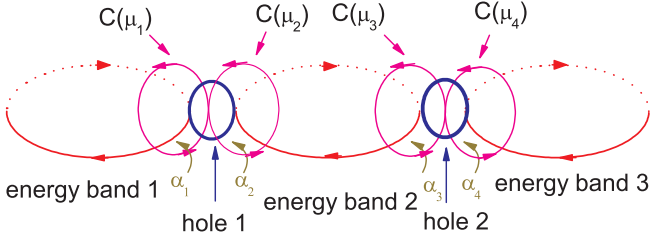}
\end{center}
\caption{(Color online) The RS of the Bloch function. $C\left(  \mu
_{j}\right)  $ is a loop formed by the trace of the zero point of $\Psi
_{L_{y}-1}\left(  z\right)  $. The energy bands are shown by closed loops. The
corresponding winding numbers are $I\left(  C\left(  \mu_{j}\right)  \right)
$=$1$ for all $j$.}%
\end{figure}

As pointed out by Hatsugai \cite{Hatsugai1, Hatsugai3}, when the Fermi energy
$\epsilon_{F}$ of the two-dimensional system lies in the $i$th energy gap, the
Hall conductance is given by the winding number of the edge state, which is
given by the number of intersections $I\left(  \alpha_{i},C\left(  \mu
_{i}\right)  \right)  $ ($\equiv I\left(  C\left(  \mu_{i}\right)  \right)  $)
between the canonical loop $\alpha_{i}$ on the RS and the trace of $\mu_{i}$.
In the present model, because there are two edge states, and, correspondingly,
there are two canonical loops in one energy gap, the Hall conductance by its
definition can be written as
\begin{equation}
\sigma_{xy}^{j,\text{edge}}=\left\{
\begin{array}
[c]{c}%
-\frac{e^{2}}{h}I\left(  C\left(  \mu_{2j-1}\right)  \right)  ,\text{
}\epsilon_{F}\leq\epsilon_{T_{j}}\\
-\frac{e^{2}}{h}I\left(  C\left(  \mu_{2j}\right)  \right)  ,\text{
\ \ }\epsilon_{F}>\epsilon_{T_{j}}%
\end{array}
\right.  \text{ }, \label{sigma}%
\end{equation}
where $\epsilon_{T_{j}}$ is the energy at the tangent point in the $j$th
energy gap. Similarly, this expression can be obtained by the Byers and Yang's
formula \cite{Byers}: Suppose that one increases an external magnetic flux
$\Phi$ from $0$ to $1$ adiabatically. According to the Laughlin-Halperin
argument \cite{Laughlin, Halperin}, when the Fermi energy lies in the $j$th
energy gap and $\epsilon_{F}\leq\epsilon_{T_{j}}$ [or $\epsilon_{F}%
>\epsilon_{T_{j}}$], $I\left(  C\left(  \mu_{2j-1}\right)  \right)  $ [or
$I\left(  C\left(  \mu_{2j}\right)  \right)  $] states are carried from the
down edge ($y=1$) to the up edge ($y=L_{y}-1$) in net. The energy change
during the adiabatic process is $\Delta E$=$I\left(  C\left(  \mu
_{2j-1}\right)  \right)  \left(  -e\right)  V_{y}$ [or $\Delta E$=$I\left(
C\left(  \mu_{2j}\right)  \right)  \left(  -e\right)  V_{y}$], where $V_{y}$
is a voltage in the $y$ direction. This gives the Hall current $I_{x}$ as
follows
\begin{equation}
I_{x}=c\frac{\Delta E}{\Phi_{0}\Delta\Phi}=\sigma_{xy}V_{y}, \label{Byer}%
\end{equation}
where $\Phi_{0}$=$hc/e$ is the flux quantum. Then we get an expression for
$\sigma_{xy}^{\text{edge}}$\ as Eq. (\ref{sigma}).

On the genus $g$=$2$ RS, the first homotopy group is generated by $4g$=$8$
generators, $\alpha_{i}$ and $\beta_{i}$, $i$=$1,\cdots,4$. We can observe
that $\mu_{2j-1}$ ($\mu_{2j}$) moves \emph{one} time across the $j$th hole,
that means $C\left(  \mu_{2j-1}\right)  $ [$C\left(  \mu_{2j}\right)  $]
$\approx\beta_{j}$ and $|I\left(  C\left(  \mu_{2j-1}\right)  \right)  |$
[$|I\left(  C\left(  \mu_{2j}\right)  \right)  |$] =$1$. Considering winding
direction (See Fig. 5 in Ref. \cite{Hatsugai1}), one can obtain that $I\left(
C\left(  \mu_{i}\right)  \right)  $=$1$ when $\phi\in\left(  0,\pi\right)  $,
while $I\left(  C\left(  \mu_{i}\right)  \right)  $=$-1$ when $\phi\in\left(
-\pi,0\right)  $ for all $i$. So, $I\left(  C\left(  \mu_{i}\right)  \right)
$=sgn$\left(  \sin\phi\right)  $, and
\begin{equation}
\sigma_{xy}^{1,\text{edge}}=\sigma_{xy}^{2,\text{edge}}=-\frac{e^{2}}%
{h}\text{sgn}\left(  \sin\phi\right)  . \label{x}%
\end{equation}

Now we turn back to make an analysis of Fig. 2 with the help of the above
results. At $\phi$=$0$, the lower energy band becomes dispersionless (Fig.
2(a)), which reflects the fact that the 2D kagom\'{e} lattice is a line graph
of the honeycomb structure \cite{Mielke}. This flat band touches at $k_{x}%
$=$0$ with the middle band, while the middle band touches at $k_{x}$%
=$\frac{2\pi}{3}$, $\frac{4\pi}{3}$ with the upper band. So there are no bulk
energy gaps and the Hall conductance is zero in this case. At $\phi\neq0,\pi$
the 2D kagom\'{e} lattice has spin chirality, and there occur two bulk band
gaps, as shown in Fig. 2(b) for $\phi$=$\pi/3$ and Fig. 2(c) for $\phi$%
=$\pi/2$. Then according to Eq. (\ref{x}), one can obtain that when the Fermi
energy lies in the bulk gaps, $\sigma_{xy}^{1,\text{edge}}$=$\sigma
_{xy}^{2,\text{edge}}$=$-\frac{e^{2}}{h}$. Note that in the case of $\phi
$=$\frac{\pi}{2}$, the middle energy band becomes flat (Fig. 2(c)) due to the
particle-hole symmetry.

Finally let us compare $\sigma_{xy}^{\text{edge}}$ (Eq. (\ref{x})) in the
present model with that in the bulk theory \cite{Ohgushi, Wang2007}. In the
latter, the bulk Hall conductance has been derived to be $\sigma
_{xy}^{1,\text{band}}$=$-\frac{e^{2}}{h}$sgn$(\sin\phi)$, $\sigma
_{xy}^{2,\text{band}}$=$0$, and $\sigma_{xy}^{3,\text{band}}$=$\frac{e^{2}}%
{h}$sgn$(\sin\phi)$ for the three bands. So, when the Fermi energy
$\epsilon_{F}$ lies in the $i$th energy gap, the bulk Hall conductance
$\sigma_{xy}^{i,\text{bulk}}$ is given by
\begin{equation}
\sigma_{xy}^{i,\text{bulk}}=\sum_{j=1}^{i}\sigma_{xy}^{j,\text{band}}%
=-\frac{e^{2}}{h}\text{sgn}(\sin\phi), \label{bulk1}%
\end{equation}
where $i$=$1,2$. Comparing Eqs. (\ref{bulk1}) with Eq. (\ref{x}), one can
obtain that
\begin{equation}
\sigma_{xy}^{\text{edge}}=\sigma_{xy}^{\text{bulk}}. \label{bulk}%
\end{equation}
This conclusion keeps consistent with the recently established common
recognition on the Hall conductance in the systems with and without edges.

In summary, we have investigated the effect of the chiral edge states on the
quantum Hall conductance in the 2D kagom\'{e} lattice with edges. According to
our derived Harper equation, there are two edge states lying in each energy
gap. They are tangent at one point in the gap, thus showing a
\textquotedblleft$\infty$\textquotedblright\ structure. The energy loops for
these two edge states move \emph{across} the holes in the RS. We have also
analyzed the winding number of these two edge states, which gives the quantum
Hall conductance $\sigma_{xy}^{\text{edge}}$=$-\frac{e^{2}}{h}$sgn$\left(
\sin\phi\right)  $ when the Fermi energy lies in the bulk gap. This conclusion
keeps consistent with that based on the topological bulk Chern-number theory.

This work was supported by NSFC under Grants Nos. 10604010 and 60776063.

\end{document}